\documentclass[twocolumn,trackchanges]{aastex701}

\usepackage{xcolor}

\usepackage{xspace}
\usepackage{amsmath} 
\usepackage[T1]{fontenc}

\usepackage[utf8]{inputenc}
\usepackage{booktabs}

\begin{document}

\title{ASASSN-15af: An Eclipsing Cataclysmic Variable above the Period Gap Showing Both Positive and Negative Superhump-like Signals}

\author[orcid=0009-0001-7046-0446,sname='Ogunwale']{Ogunwale Bisi Bernard}
\affiliation{Department of Physics, Ariel University, Ariel 40700, Israel}
\email[show]{tok4bisiben@gmail.com}

\author[orcid=0000-0003-0155-2539, sname='Michael']{Michael Shara} 
\affiliation{Department of Astrophysics, American Museum of Natural History, New York, NY 10024, USA}
\email{mshara@amnh.org}

\author[orcid=0000-0002-1836-3120 , sname='Shporer']{Avi Shporer}
\affiliation{Kavli Institute for Astrophysics and Space Research, Massachusetts Institute of Technology, Cambridge, MA 02139, USA}
\email{shporer@mit.edu}

\author[orcid=0000-0002-7349-1109, sname='Guetta']{Dafne Guetta} 
\affiliation{Department of Physics, Ariel University, Ariel 40700, Israel}
\affiliation{Astrophysics, Geophysics, and Space Science Research Center, Ariel University, Ariel 40700, Israel}
\email{dafneg@ariel.ac.il}

\author[0000-0003-3757-1440,sname='Tal-Or']{Lev Tal-Or}
\affiliation{Department of Physics, Ariel University, Ariel 40700, Israel}
\affiliation{Astrophysics, Geophysics, and Space Science Research Center, Ariel University, Ariel 40700, Israel}
\email[show]{levtalor@ariel.ac.il}

\begin{abstract}
We present a photometric study of ASASSN-15af (TIC~231050794), combining \textit{TESS} full-frame-image photometry from the TEQUILA project with 2-minute SPOC light curves and archival ASAS-SN and ATLAS monitoring over 4400~d. 
The long-term light curve shows recurrent dwarf-nova outbursts, establishing ASASSN-15af as a high-inclination eclipsing dwarf nova above the canonical period gap. From 391 quiescent eclipses, we derive the working ephemeris $T(E)=2139.76108(3)+0.14982063(2)\,E$ (BTJD), corresponding to $P_{\rm orb}=3.596$~hr. A quadratic fit is formally better, but the timings sample only three widely separated epochs, so we regard the apparent O--C curvature as tentative. In quiescence, the out-of-eclipse waveform is dominated by a stable $2f_{\rm orb}$ component, consistent with a substantial ellipsoidal contribution from the donor; during the Sector~96 outburst, its fractional amplitude decreases by $\simeq66\%$, consistent with dilution by the brightened disk. The outburst-only signals occur at periods $\sim1.6\%$ shorter than $P_{\rm orb}$ in Sector~2 and $\sim3.1\%$ longer in Sector~96, consistent with negative- and positive-superhump-like modulations, respectively. Because each is detected in only one outburst, their recurrence remains to be established. Their period-offset ratio, $\phi\simeq0.51$, is consistent with empirical trends among systems showing both modes, while the inferred superhump-based mass ratios span $q\simeq0.06$--$0.24$. During the Sector~96 outburst, the eclipse deepens by $\Delta D=0.196\pm0.028$ and its centroid shifts earlier by $\Delta t_{\rm c}=-66.6\pm7.3$~s, indicating a redistribution of accretion-powered light. These results identify ASASSN-15af as an above-gap eclipsing dwarf nova with a dynamically variable accretion disk; spectroscopy is needed to constrain the mass ratio.
\end{abstract}

\section{Introduction}
\label{sec:intro}

Cataclysmic variables (CVs) are interacting compact binaries in which the primary star, a white dwarf, accretes matter from a Roche-lobe-filling, late-type, low-mass companion, usually an M-type star \citep{warner_1995,knigge_2011}. In most non-magnetic systems, the transferred material forms an accretion disk around the white dwarf, while the impact of the stream with the outer disk produces a bright spot. The interplay among mass transfer, disk accretion, and binary geometry gives rise to rich variability on timescales ranging from seconds to years, making CVs important laboratories for studying accretion physics, close-binary evolution, and angular-momentum loss in compact systems \citep{warner_1995,knigge_2011}. Their orbital-period distribution is marked by the well-known period gap near 2--3 hr, which separates the short-period and long-period CV populations and plays a central role in the standard evolutionary picture \citep{warner_1995,knigge_2011}.

Eclipsing CVs are particularly important because their high orbital inclinations provide direct geometric constraints on the binary components and the accretion flow \citep{wood_1986,horne_1985,littlefair_2008,savoury_2011}. The eclipse profile encodes information on the white dwarf, accretion disk, and bright spot through its depth, duration, and detailed ingress and egress morphology; combined with time-resolved spectroscopy, eclipse modeling can yield precise system parameters including the orbital inclination, mass ratio, and component masses \citep{littlefair_2008,savoury_2011}.  Even from photometry alone, eclipse timing provides a precise orbital clock, while the out-of-eclipse (OOE) waveform and its response to accretion-state changes probe the evolving structure of the disk and bright spot.  High-inclination CVs above the period gap comprise a heterogeneous population that includes dwarf novae, nova-like systems, and SW~Sex stars. The SW~Sex subclass is concentrated primarily in the $\sim3$--4.5~hr period range and is defined mainly by spectroscopic characteristics, including predominantly single-peaked emission lines, transient absorption near orbital phase 0.5, and phase offsets between photometric and spectroscopic ephemerides \citep{thorstensen_1991,rodriguez-gil_2007b,schmidtobreick_2013}. These characteristics make SW~Sex systems a relevant comparison class for some above-gap eclipsing CVs, but orbital period and inclination alone are not sufficient for an SW~Sex classification.

The advent of space-based time-domain photometry has significantly advanced the study of eclipsing CVs. In particular, the Transiting Exoplanet Survey Satellite \citep[TESS;][]{ricker_2015} provides long, nearly continuous light curves with cadences well suited to tracing orbital modulation, eclipse morphology, and changes in accretion state. For many accreting binaries, \textit{TESS} data enable precise determinations of orbital periods and eclipse ephemerides, the construction of high signal-to-noise phase-folded eclipse templates, and comparisons of eclipse properties across different luminosity states \citep[e.g.,][]{bruch_2022}. 

ASASSN-15af (TIC~231050794; 2MASS~J01575499$-$5430367) was identified as a candidate eclipsing compact binary in the TEQUILA project \citep{ogunwale_2025}, a \textit{TESS} full-frame-image survey of variable sources. The source had previously been associated in catalog compilations with the unrelated H$\alpha$-emitting Orion T~Tauri star 2MASS~J05181685$-$0537300, which was studied by \citet{wiramihardja_1991} and \citet{lee_2007}. Those studies did not concern ASASSN-15af itself; rather, the young-star classification appears to have propagated through an erroneous catalog cross-match. \citet{schmidt_2019} inherited this association and consequently excluded ASASSN-15af from their confirmed M-dwarf flare sample. The two sources are clearly distinct: TIC~231050794 is located at RA~=~01:57:55, Dec~=~$-$54:30:36 ($b\simeq-60\degr$), whereas the Orion object lies at a completely different sky position. We therefore treat the earlier T~Tauri association as a catalog-identification error.

More recent studies have instead placed ASASSN-15af in compact-binary contexts using Gaia~DR3 \citep{gaia_2023}. \citet{canbay_2023} included the source in their Gaia~DR3-based catalog as a non-magnetic CV candidate, while \citet{shani_2025} independently identified it as a post-common-envelope white-dwarf--main-sequence (WDMS) binary with $P_{\rm orb}=0.149823$~d from a \textit{TESS}-based periodicity search, although it was left unclassified within their eclipsing WDMS sample. Their reported period agrees well with our refined ephemeris (Section~\ref{subsec:eclipse_timing}).

In this paper, we present a photometric study of ASASSN-15af, primarily based on \textit{TESS} observations. Our goals are to refine the orbital period and eclipse ephemeris, characterize the eclipse morphology across accretion states, constrain the binary mass ratio from superhump-like signals, and place the system in the context of above-gap eclipsing dwarf novae. Together, these results provide the first quantitative photometric framework for future follow-up studies, including high-speed multi-band eclipse photometry and time-resolved spectroscopy needed to confirm the classification and determine the system parameters in greater detail.

The paper is organized as follows. In Section~\ref{sec:data}, we describe the observations and data preparation. In Section~\ref{sec:period_morph}, we present the period analysis, eclipse-template construction, and timing analysis. In Section~\ref{sec:results}, we examine the state-dependent photometric variability, including the orbital-harmonic structure, superhump-like signals, eclipse morphology, and broadband spectral properties. In Section~\ref{sec:discussion}, we discuss the classification and astrophysical implications of ASASSN-15af in the broader context of eclipsing CVs, and we present our conclusions in Section~\ref{sec:conclusion}.

\begin{figure*}[t!]
    \centering
    \includegraphics[width=0.95\linewidth]{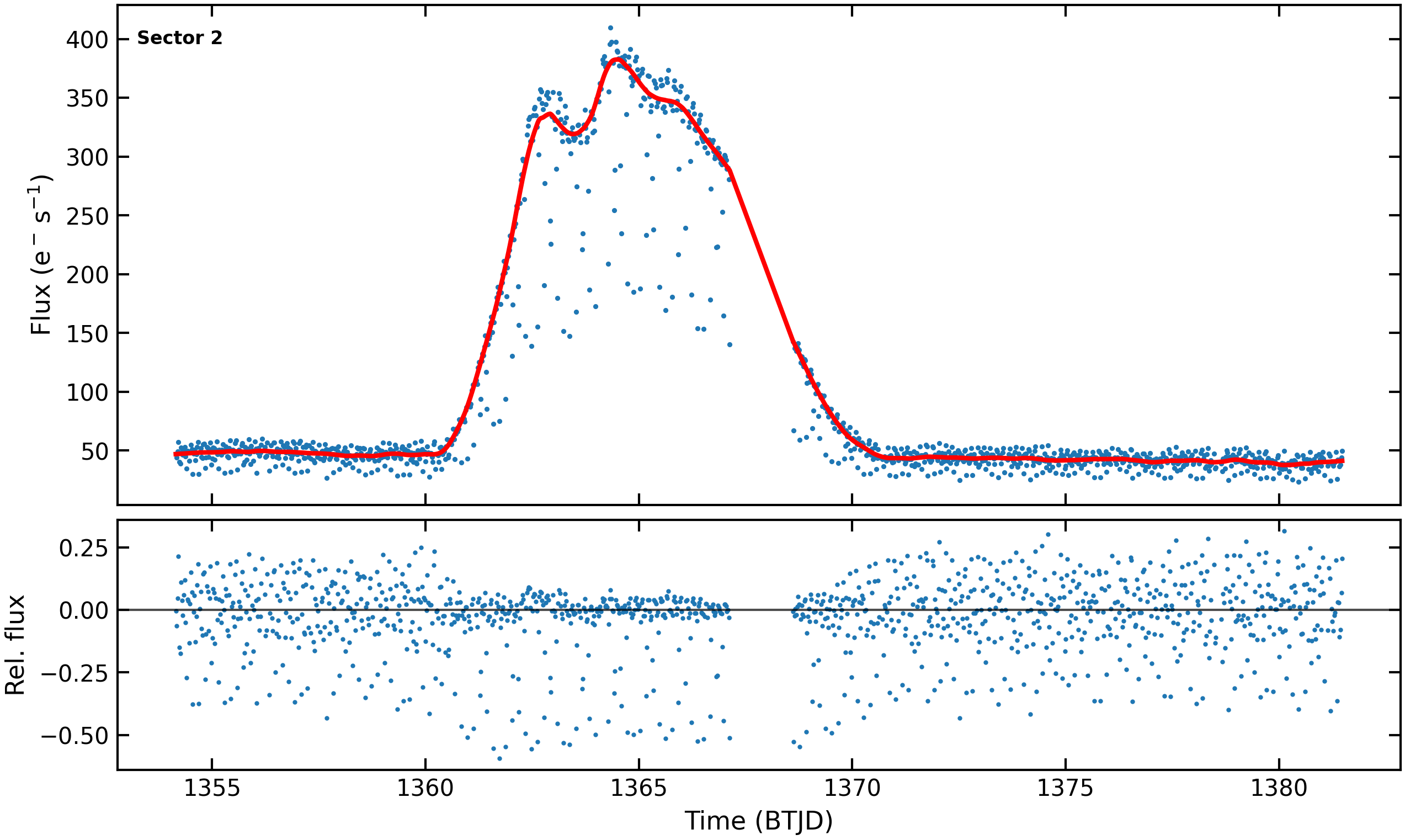}
    \caption{TEQUILA full-frame-image light curve of ASASSN-15af obtained in \textit{TESS} Sector~2. The upper panel shows the observed flux together with the fitted LOWESS trend, while the lower panel shows the corresponding detrended normalized light curve, plotted as $F/F_{\rm trend}-1$. The light curve reveals a $\sim10$~day outburst event, together with recurring eclipses, which first drew attention to the system as a candidate eclipsing cataclysmic variable.}
    \label{fig:tequila_lc}
\end{figure*}

\section{TESS Observations and Archival Data Sources}
\label{sec:data}

In this section, we describe the observations of ASASSN-15af and the data products used in our analysis. Our study is based primarily on the TESS data, including full-frame-image photometry from the TEQUILA project \citep{ogunwale_2025} and 2-minute cadence light curves produced by the Science Processing Operations Center \citep[SPOC;][]{jenkins_2016}. To place the \textit{TESS} behavior in a broader temporal context, we also examined archival ground-based photometry from the All-Sky Automated Survey for Supernovae \citep[ASAS-SN;][]{shappee_2014,kochanek_2017} and the Asteroid Terrestrial-impact Last Alert System \citep[ATLAS;][]{tonry_2018,heinze_2018}. Basic archival broadband photometric properties of ASASSN-15af are summarized in Table~\ref{tab:archival_phot}.

\begin{deluxetable}{lcc}
\tablecaption{Archival Broad-band Photometry of ASASSN-15af\label{tab:archival_phot}}
\tablehead{
\colhead{Band} & \colhead{Magnitude} & \colhead{Uncertainty}
}
\startdata
TESS       & 16.3185 & 0.0208 \\
$B$        & 17.2090 & 0.1170 \\
$V$        & 17.2126 & 0.0664 \\
Gaia $G$   & 16.9334 & 0.0199 \\
$J$        & 15.1250 & 0.0400 \\
$H$        & 14.6300 & 0.0380 \\
$K_s$      & 14.2660 & 0.0660 \\
WISE W1    & 14.1460 & 0.0260 \\
WISE W2    & 13.9190 & 0.0340 \\
\enddata
\tablecomments{
Archival catalog magnitudes of ASASSN-15af compiled from TIC v8.2 as listed on the TFOP/\textit{TESS} target pages. The $B$ and $V$ values are compiled or derived catalog estimates.}
\end{deluxetable}

\subsection{TESS observations}

\textit{TESS} data are available through two main photometric pathways: short-cadence target observations and full-frame images (FFIs). For selected targets, SPOC provides calibrated light curves and target-pixel products, including Simple Aperture Photometry (SAP) and Pre-search Data Conditioning SAP (PDCSAP). In parallel, the FFIs enable the extraction of light curves for a much larger population of sources beyond the pre-selected target list. This distinction is particularly important for identifying and characterizing variable objects whose nature may not have been recognized at the time of target selection.

The TEQUILA project \citep{ogunwale_2025} was developed to exploit this discovery space by producing differential-photometry light curves for variable sources in the \textit{TESS} FFIs over the first two years of the mission. By working directly with FFI-based photometry, TEQUILA conducted a blind search for variable sources and enabled the identification of objects that might otherwise have gone unrecognized or poorly classified.

\begin{figure*}[t!]
    \centering
    \includegraphics[width=0.95\linewidth]{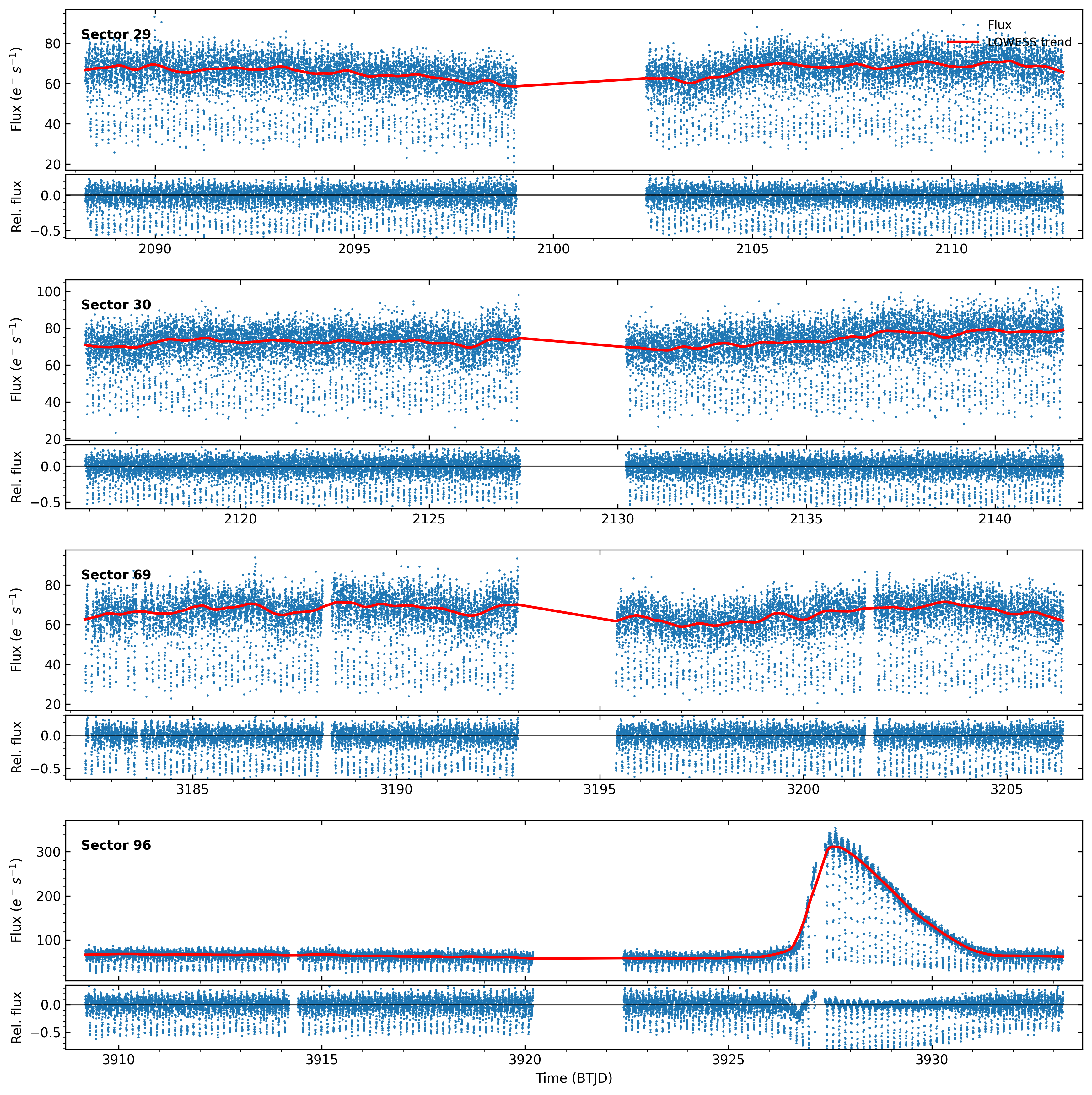}
    \caption{
    SPOC 2-minute SAP light curves of ASASSN-15af obtained in \textit{TESS} Sectors~29, 30, 69, and 96. In each sector, the upper panel shows the SAP flux and fitted LOWESS trend (red line), and the lower panel shows the detrended light curve normalized as $F/F_{\rm trend}-1$. Sector~96 additionally contains a shorter-duration outburst, distinct from the longer event seen in the TEQUILA Sector~2 light curve.
    }
    \label{fig:spoc_all_sectors}
\end{figure*}

ASASSN-15af corresponds to TIC~231050794 in the TESS Input Catalog \citep{stassun_2019}, at RA~=~01:57:55.03, Dec~=~$-$54:30:36.35 (J2000). A $\sim10$~day brightening event was identified in the TEQUILA Sector~2 light curve (Figure~\ref{fig:tequila_lc}), drawing attention to the source as an object of interest. In the extended \textit{TESS} mission, it was later observed at 2-minute cadence in several sectors, and SPOC delivered calibrated light curves for those observations. In this work, we analyze the available SPOC light curves from Sectors~29, 30, 69, and 96 together with the TEQUILA light curve from Sector~2. The adopted SPOC SAP light curves are shown in Figure~\ref{fig:spoc_all_sectors}. These data sample the source primarily in quiescence, except for Sector~96, which contains a shorter-duration brightening event. All times used in this work are expressed in Barycentric TESS Julian Date (BTJD), defined as $\mathrm{BTJD}=\mathrm{BJD}_{\mathrm{TDB}}-2457000.0$, where TDB denotes Barycentric Dynamical Time.

For the SPOC data, we adopted SAP fluxes instead of PDCSAP fluxes because our primary goal is to preserve the source's intrinsic accretion-driven variability and eclipse morphology. In accreting binaries, detrending procedures optimized for exoplanet searches can partially suppress or distort astrophysical variability on timescales relevant to accretion-state changes and outburst evolution. We therefore use the SAP light curves as the starting point for our timing and morphology analysis, and apply our own detrending procedure as described below. All data points with \texttt{QUALITY} flag other than zero were removed from further analysis.

\subsection{Archival ground-based photometry}
\label{subsec:archival_data}

The ASAS-SN data provide long-term $V$- and $g$-band monitoring spanning approximately 4400~d. The light curve was obtained by combining the archival survey export with a more recent Sky Patrol download \citep{hart_2023}, retaining only detections brighter than the limiting magnitude with finite uncertainties satisfying $\sigma_{\rm mag}<1.0$~mag and quality flag \texttt{G}. For ATLAS, we used forced photometry at the position of ASASSN-15af and retained only $o$-band detections with positive flux, signal-to-noise ratio $\geq2.5$, magnitude uncertainty $\leq0.5$~mag, and reduced chi-squared $\chi^2_\nu\leq3.0$; the $c$-band data were excluded owing to their sparser sampling and higher rate of spurious detections at this source brightness.

\begin{figure*}[t!]
    \centering
    \includegraphics[width=0.95\linewidth]{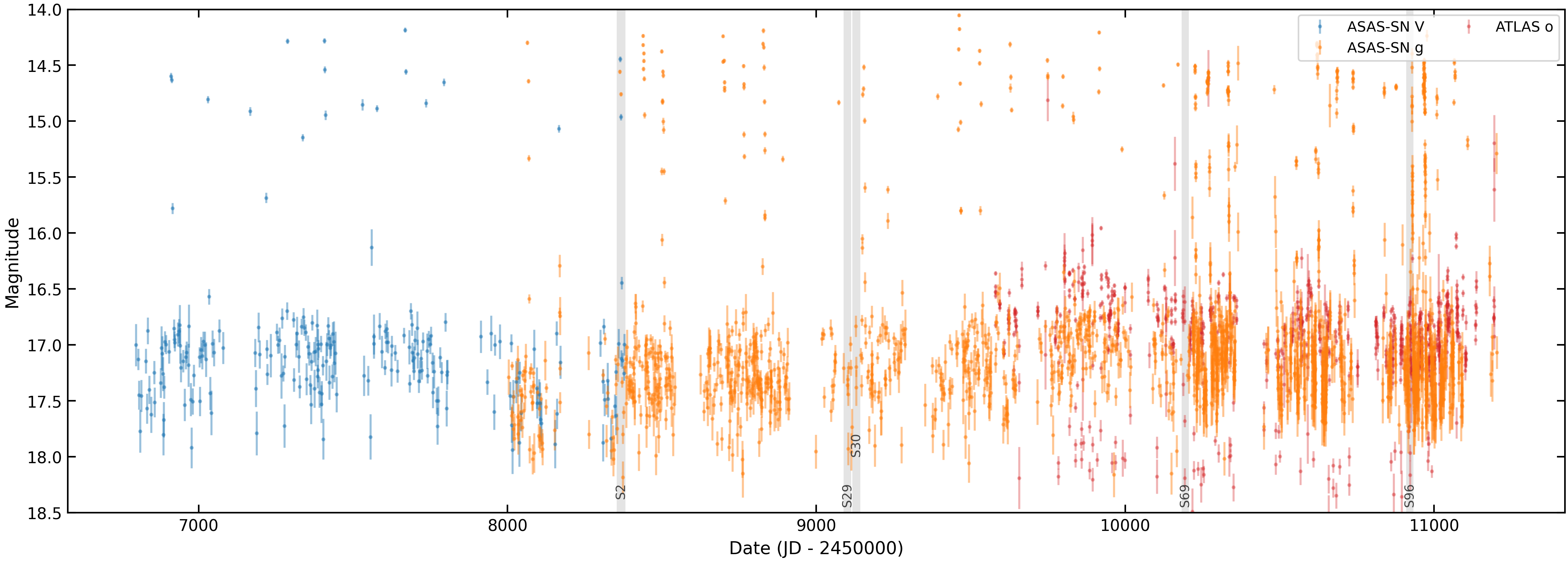}
    \caption{
    Long-term optical light curve of ASASSN-15af from archival ASAS-SN ($V$ and $g$) and ATLAS ($o$) photometry, with the \textit{TESS} observing windows shaded. The source spends most epochs near a faint baseline around $g\sim17$--18 mag, but undergoes repeated brightenings to $\sim14$--16 mag over the archival baseline. The two \textit{TESS} outbursts are therefore part of a broader pattern of recurrent accretion-driven variability.
    }
    \label{fig:longterm_archival}
\end{figure*}

Figure~\ref{fig:longterm_archival} shows the combined ASAS-SN and ATLAS light curve, with the \textit{TESS} observing windows marked for reference. The archival data reveal that the Sector~2 and Sector~96 brightenings are not isolated episodes but part of a pattern of recurrent outbursts over the full baseline. We identify 50 discrete high-state epochs (detections brighter than $g\simeq16.0$) over the 4400~d ASAS-SN baseline, with a median recurrence interval of approximately 69~d (range 35--273~d) and an observed duty cycle of $\sim$5.6\%; the latter should be regarded as a lower limit given the sparse cadence of the ground-based monitoring. No extended faint state is detected over the full baseline. The source otherwise resides near a quiescent level of $g\sim17$--18~mag between outbursts.

\section{Orbital Period and Eclipse Morphology}
\label{sec:period_morph}

In this section, we determine the orbital period of ASASSN-15af and quantify the basic properties of its eclipse morphology. Prior to the period analysis, we removed long-timescale variations in the out-of-eclipse flux using the locally weighted regression  \citep[LOWESS;][]{cleveland_1979}. The adopted smoothing span was set to 0.0323, corresponding to approximately five orbital cycles, and three iterations were performed to reduce the influence of eclipse points and other local deviations. The detrending was performed by dividing each SAP flux light curve by its LOWESS trend, thereby suppressing slow secular variations.

The upper panels of Figure~\ref{fig:spoc_all_sectors} show the original SAP light curves together with the fitted LOWESS trends, while the lower panels show the corresponding detrended light curves used for the period and morphology analysis. We searched for the orbital periodicity using Box Least Squares \citep[BLS;][]{kovacs_2002} and phase-dispersion minimization \citep[PDM;][]{stellingwerf_1978} applied to the detrended SPOC light curves. The strongest BLS peak occurs at $P = 0.149824$~d, while the deepest PDM minimum occurs at $P = 0.149819(42)$~d, where the uncertainty was estimated by fitting a parabola around the minimum following \citet{fernie_1989}. The close agreement between the two methods confirms the orbital periodicity near 0.14982~d. A secondary signal near 0.0749~d is interpreted as the first harmonic of the strongly non-sinusoidal eclipse waveform. 

\begin{figure}
    \centering
    \includegraphics[width=0.48\textwidth, height=5.0cm]{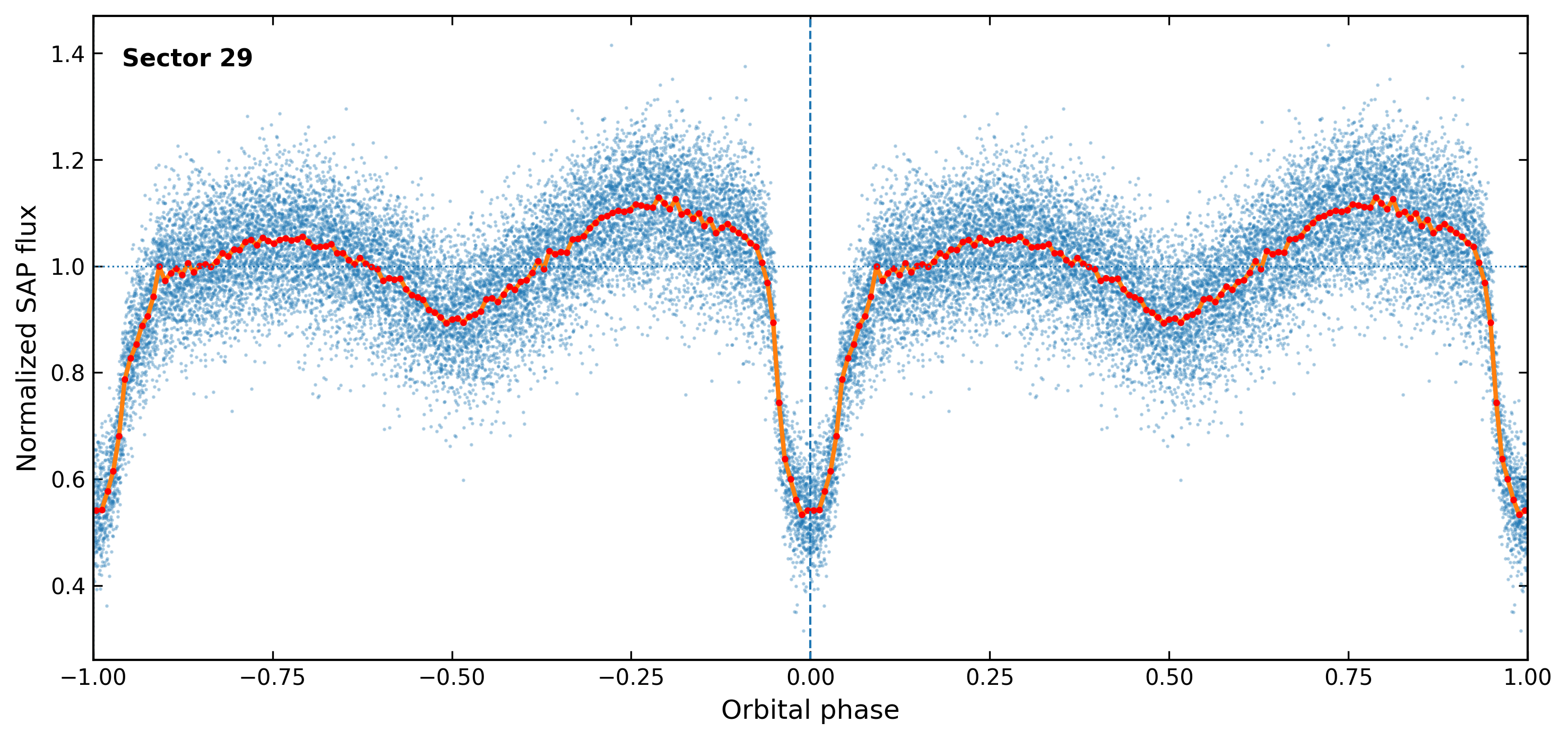}
    \caption{
    Phase-folded Sector~29 SAP light curve of ASASSN-15af using the working orbital period derived from the period search. Two cycles are shown for clarity, with the primary eclipse centered at phase 0. The coherent repetition of the eclipse profile supports the adopted orbital solution.
    }
    \label{fig:phase_folded_lc}
\end{figure}

Using the orbital period above, we phase-folded the Sector~29 \textit{TESS} light curve, adopting $T_0=2088.3738$~BTJD as a preliminary reference epoch solely to define orbital phase; we derive the refined eclipse-timing ephemeris below. The resulting profile (Figure~\ref{fig:phase_folded_lc}) shows a deep primary eclipse and pronounced out-of-eclipse modulation, including a broad minimum near phase~0.5 and enhanced flux before eclipse, consistent with a high-inclination accreting binary. Folding on the shorter $\sim0.0749$~d signal instead merges alternating orbital cycles, confirming that it represents the first harmonic of the orbital waveform.

\subsection{Eclipse timings and refined ephemeris}
\label{subsec:eclipse_timing}

We characterized the mean eclipse morphology of ASASSN-15af using the detrended SPOC light curves from Sectors~29, 30, and 69. Individual eclipses were locally normalized and combined to construct sector-specific empirical templates. The central portion of each template was interpolated using a piecewise cubic Hermite interpolating polynomial \citep[PCHIP;][]{fritsch_1984} and matched to individual eclipses, allowing the eclipse depth and mid-eclipse time to vary. After quality filtering, we obtained 391 mid-eclipse times spanning Sectors~29 ($E=-343$ to $-182$), 30 ($E=-161$ to $+10$), and 69 ($E=+6957$ to $+7119$), with median timing uncertainties of approximately 24, 29, and 19~s, respectively.

We fitted these timings with a linear ephemeris,
\begin{equation}
T(E)=T_0+P_{\rm orb}E,
\end{equation}
including an additional timing-jitter term added in quadrature to the individual timing uncertainties. The fit gives a mean orbital period of
\[
P_{\rm orb}=0.1498206293~{\rm d},
\]
with a fitted timing jitter of $\simeq27.5$~s and a formal statistical period uncertainty of $6.1\times10^{-9}$~d. Because the individual timings are concentrated in only three widely separated \textit{TESS} observing epochs, the formal statistical uncertainty does not fully capture the sensitivity of the period to epoch-dependent timing offsets. Fits to the two independent long-baseline combinations, Sectors~29+69 and 30+69, differ from the full-data solution by $1.4$--$1.6\times10^{-8}$~d. Accounting for this sensitivity, we adopt the more conservative period uncertainty of $2\times10^{-8}$~d and use
\begin{equation}
T(E)=2139.76108(3)+0.14982063(2)\,E
\quad {\rm (BTJD)}
\label{eq:linear_result}
\end{equation}
as the working photometric ephemeris.

The eclipse timings retain an rms O--C scatter of $\simeq37$~s, substantially larger than the individual formal uncertainties. A quadratic ephemeris provides a formally improved fit and yields
\[
\frac{dP}{dE}=(2.10\pm0.63)\times10^{-10}\ {\rm d~cycle^{-1}},
\]
but the timings sample only three widely separated epochs and remain affected by intrinsic timing variability. We therefore regard the quadratic term as tentative evidence for O--C curvature, and retain the linear ephemeris of Equation~\ref{eq:linear_result} for all subsequent phase-folding and morphology analyses.

\section{State-dependent Photometric Variability}
\label{sec:results}

Having established the orbital period and eclipse properties of ASASSN-15af in Section~\ref{sec:period_morph}, we now examine how its photometric behavior changes between quiescence and outburst. We focus on the evolution of the OOE waveform, the periodic signals present in each accretion state, and the corresponding changes in eclipse depth and morphology. Particular attention is given to Sectors~2 and 96, which contain well-defined outburst intervals and sufficient quiescent coverage for direct comparison.

{
\subsection{Orbital-harmonic structure and ellipsoidal modulation}
\label{subsubsec:ellipsoidal}

The OOE light curves show a persistent double-wave morphology (Figure~\ref{fig:phase_folded_lc}), suggesting that a substantial fraction of the orbital modulation occurs at $2f_{\rm orb}$. Such a signal is expected from ellipsoidal modulation of a tidally distorted donor, which produces two maxima and two minima per orbital cycle, whereas irradiation-dominated variability is generally strongest at the orbital fundamental \citep[e.g.,][]{morris_1985,morris_naftilan_1993}.

To quantify the harmonic content, we masked the primary eclipse ($|\phi|<0.07$) and fitted each detrended OOE light curve with
\begin{equation}
F(\phi)=
C+\sum_{k=1}^{3}
\left[
C_k\cos(2\pi k\phi)+
S_k\sin(2\pi k\phi)
\right],
\end{equation}
where $k=1$, 2, and 3 correspond to $f_{\rm orb}$, $2f_{\rm orb}$, and $3f_{\rm orb}$, respectively. The semi-amplitude of each component is
\begin{equation}
A_{kf}=\left(C_k^2+S_k^2\right)^{1/2}.
\end{equation}
Uncertainties were estimated by bootstrap resampling complete orbital cycles, thereby preserving correlated variability within each orbit. The results for the adopted 0.8-d LOWESS detrending window are summarized in Table~\ref{tab:harmonic_results}.

\begin{deluxetable*}{llccccccc}
\tablecaption{Orbital-harmonic decomposition of the out-of-eclipse light curves
\label{tab:harmonic_results}}
\tablehead{
\colhead{Sector} &
\colhead{State} &
\colhead{$A_f$} &
\colhead{$A_{2f}$} &
\colhead{$A_{3f}$} &
\colhead{$A_{2f}/A_f$} &
\colhead{$A_{3f}/A_{2f}$} &
\colhead{$\phi_{2f}$} &
\colhead{$\Delta {\rm BIC}_{2\rightarrow3}$}
}
\startdata
29 & Quiescence & $0.0374^{+0.0012}_{-0.0013}$ & $0.0745^{+0.0012}_{-0.0012}$ & $0.0033^{+0.0011}_{-0.0010}$ & $1.99^{+0.09}_{-0.08}$ & $0.045^{+0.015}_{-0.014}$ & $0.2831^{+0.0010}_{-0.0012}$ & $-8.7$ \\
30 & Quiescence & $0.0340^{+0.0012}_{-0.0012}$ & $0.0699^{+0.0013}_{-0.0012}$ & $0.0056^{+0.0011}_{-0.0010}$ & $2.06^{+0.09}_{-0.08}$ & $0.080^{+0.016}_{-0.015}$ & $0.2822^{+0.0011}_{-0.0011}$ & $9.2$ \\
69 & Quiescence & $0.0343^{+0.0016}_{-0.0015}$ & $0.0843^{+0.0012}_{-0.0013}$ & $0.0059^{+0.0010}_{-0.0010}$ & $2.46^{+0.13}_{-0.12}$ & $0.070^{+0.012}_{-0.012}$ & $0.2849^{+0.0009}_{-0.0009}$ & $21.9$ \\
96 & Quiescence & $0.0505^{+0.0015}_{-0.0017}$ & $0.0918^{+0.0016}_{-0.0016}$ & $0.0046^{+0.0011}_{-0.0011}$ & $1.82^{+0.06}_{-0.06}$ & $0.050^{+0.011}_{-0.012}$ & $0.2810^{+0.0012}_{-0.0011}$ & $0.8$ \\
96 & Outburst & $0.0497^{+0.0046}_{-0.0053}$ & $0.0313^{+0.0021}_{-0.0019}$ & $0.0071^{+0.0020}_{-0.0019}$ & $0.63^{+0.10}_{-0.08}$ & $0.226^{+0.063}_{-0.060}$ & $0.2716^{+0.0061}_{-0.0066}$ & $70.2$ \\
\enddata
\tablecomments{The quoted semi-amplitudes and uncertainties are the median values and 68\% intervals obtained by bootstrap resampling complete orbital cycles using the adopted 0.8-d LOWESS detrending window.  We define $\Delta{\rm BIC}{2\rightarrow3}\equiv{\rm BIC}{k\leq2}-{\rm BIC}{k\leq3}$, where the two models include harmonics through $2f{\rm orb}$ and $3f_{\rm orb}$, respectively. Positive $\Delta{\rm BIC}{2\rightarrow3}$ therefore favors inclusion of the $3f{\rm orb}$ term, whereas negative values favor the model truncated at $2f_{\rm orb}$}

\end{deluxetable*}

In all four short-cadence quiescent sectors, the $2f_{\rm orb}$ component is stronger than the orbital fundamental, with $A_{2f}/A_f\simeq1.8$--$2.5$ (Table~\ref{tab:harmonic_results}). Its phase is also highly stable across the widely separated epochs, with the first maximum near $\phi_{2f}\simeq0.28$. An ideal symmetric ellipsoidal waveform peaks near quadrature, $\phi\simeq0.25$ and 0.75; the small but persistent phase offset indicates that the double-wave modulation is not purely ellipsoidal and likely includes contributions from asymmetric accretion structures. The $3f_{\rm orb}$ component remains weak in quiescence, contributing only a few percent of the $2f_{\rm orb}$ amplitude.

Sector~96 provides a direct comparison between accretion states. During outburst, the orbital fundamental remains comparable to its quiescent value, whereas the $2f_{\rm orb}$ component is strongly reduced. Independent sampling of the quiescent and outburst orbit-bootstrap distributions gives
\begin{equation}
\frac{A_{2f,\rm out}}{A_{2f,\rm qui}}
=
0.341^{+0.023}_{-0.022},
\end{equation}
corresponding to a $65.9^{+2.1}_{-2.3}\%$ reduction in the fractional $2f_{\rm orb}$ amplitude. This decrease is qualitatively consistent with dilution of a donor-related ellipsoidal modulation by the brighter accretion disk during outburst.

The harmonic structure also becomes less purely double-wave during outburst. The relative contribution of the $3f_{\rm orb}$ component increases from $A_{3f}/A_{2f}\simeq0.05$ in quiescence to $\simeq0.23$ during outburst, and inclusion of the $3f_{\rm orb}$ term is strongly favored in the outburst fit. However, the absolute increase in $A_{3f}$ is only weakly constrained; the change therefore primarily reflects the strong suppression of $2f_{\rm orb}$.

We tested the sensitivity of these results to the detrending timescale by repeating the analysis with LOWESS windows of 0.8--2.0~d. The quiescent harmonic amplitudes and phases remain essentially unchanged, and the inferred Sector~96 suppression of the $2f_{\rm orb}$ component remains within 63--66\%. The state-dependent change in harmonic structure is therefore insensitive to the adopted detrending scale.

The persistent, phase-coherent $2f_{\rm orb}$ modulation and its strong fractional suppression as the system brightens are consistent with a substantial ellipsoidal contribution from the donor. However, the non-zero fundamental, higher-order harmonic structure, and superhump-like variability discussed below show that the OOE waveform also contains state-dependent accretion-related contributions. We therefore interpret the observed modulation as a superposition of a persistent donor-related double-wave component and variable emission from the accretion flow. Because the observed $2f_{\rm orb}$ amplitude depends on the binary geometry as well as contributions from the disk, bright spot, irradiation, and other asymmetric structures, it cannot be converted directly into a unique donor light fraction.

\begin{figure*}[t!]
    \centering
    \includegraphics[width=0.98\linewidth]{%
        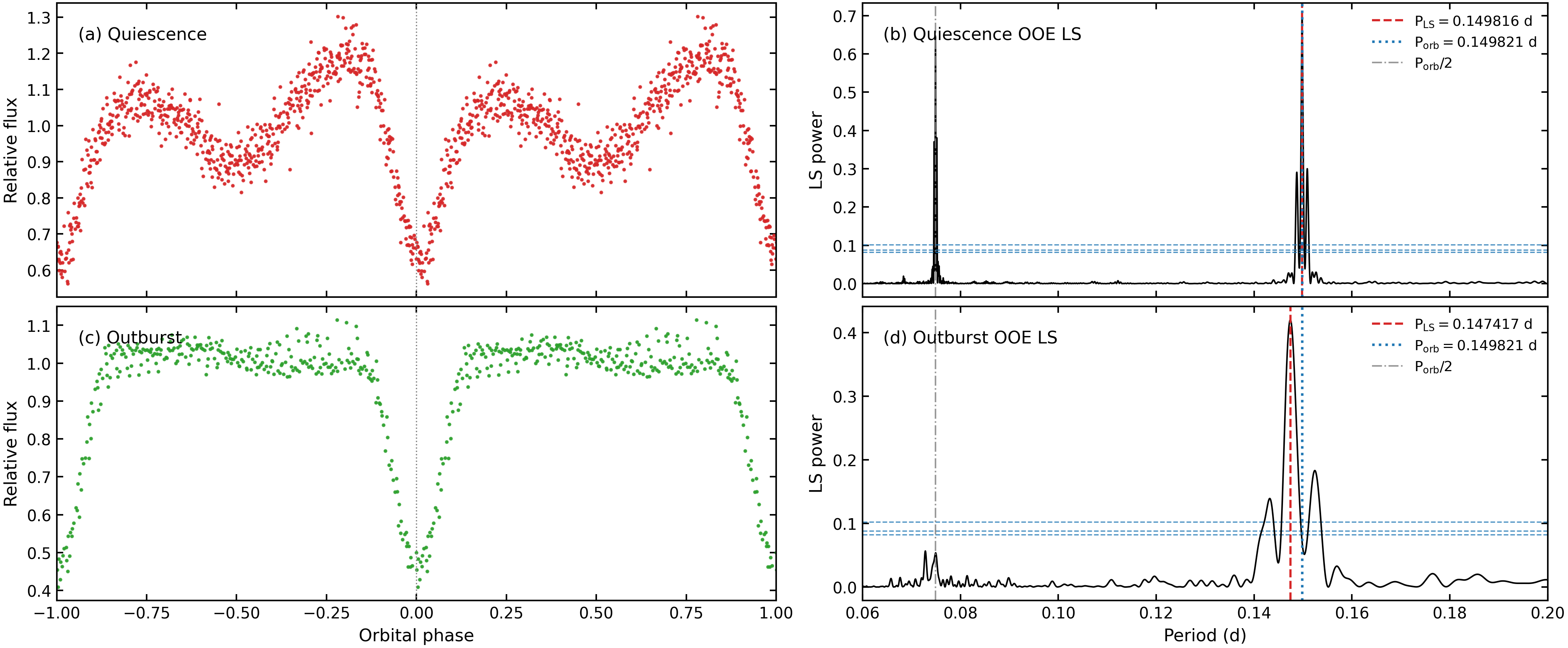}
    \caption{
    Quiescent and outburst OOE variability in Sector~2 (30-minute FFI cadence).
    \textit{Left panels:} phase-folded light curves in quiescence (top) and 
    outburst (bottom), folded on the adopted orbital period. 
    \textit{Right panels:} Lomb--Scargle periodograms of the OOE data.
    In quiescence, the dominant OOE period is consistent with the orbital 
    period. During the outburst, the dominant signal shifts to a shorter period, 
    $P_{\rm LS} = 0.147417$~d ($\sim 1.6\%$ below $P_{\rm orb}$), consistent 
    with a negative superhump-like modulation.
    }
    \label{fig:s2_state_compare}
\end{figure*}

\begin{figure*}[t!]
    \centering
    \includegraphics[width=0.98\linewidth]{%
        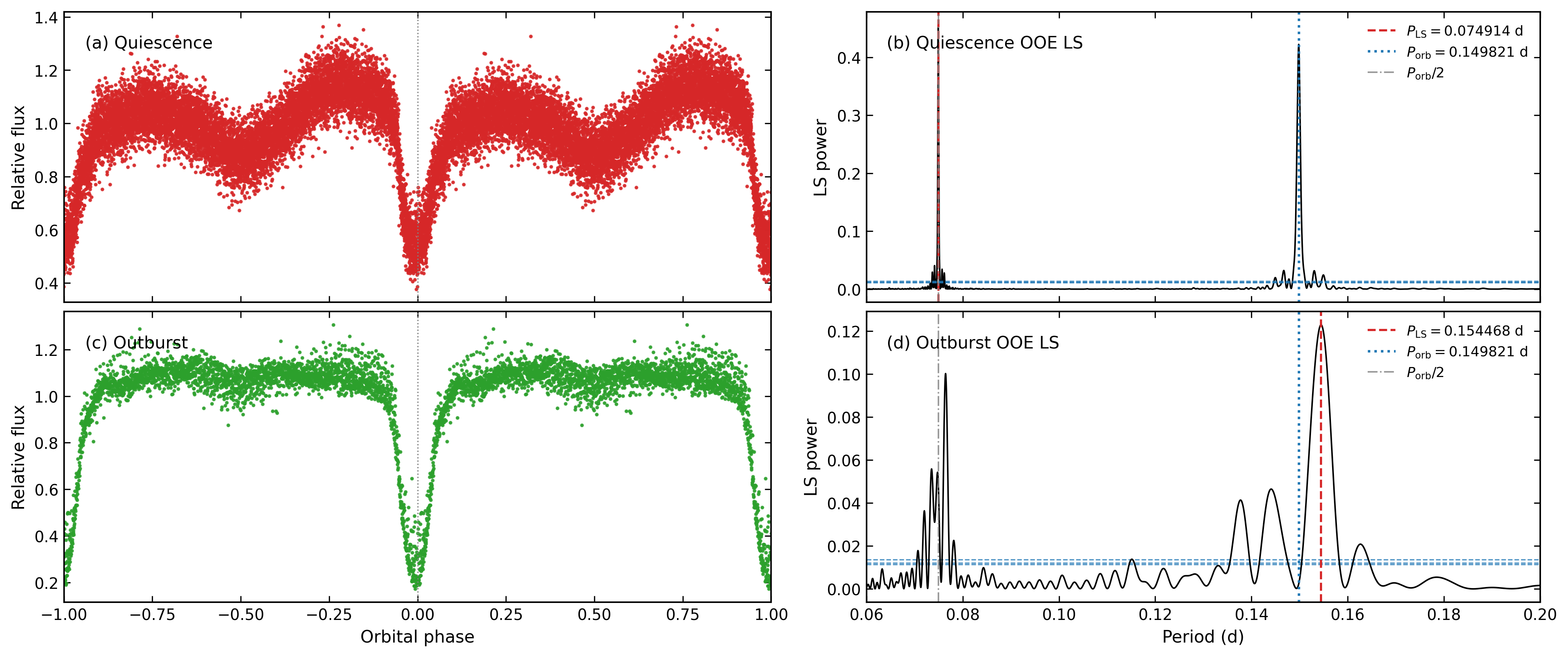}
    \caption{
    Quiescent and outburst OOE variability in Sector~96 (2-minute SPOC cadence).
    \textit{Left panels:} phase-folded light curves in quiescence (top) and 
    outburst (bottom), folded on the adopted orbital period. 
    \textit{Right panels:} Lomb--Scargle periodograms of the OOE data.
    The quiescent OOE signal is dominated by a double-humped waveform with 
    strongest power near $P_{\rm orb}/2$, whereas the outburst state is 
    characterized by a longer dominant modulation at 
    $P_{\rm LS} = 0.154468$~d ($\sim 3.1\%$ above $P_{\rm orb}$), consistent 
    with a positive superhump-like modulation.
    }
    \label{fig:s96_state_compare}
\end{figure*}

\begin{figure*}[t!]
    \centering
    \includegraphics[width=0.98\linewidth]{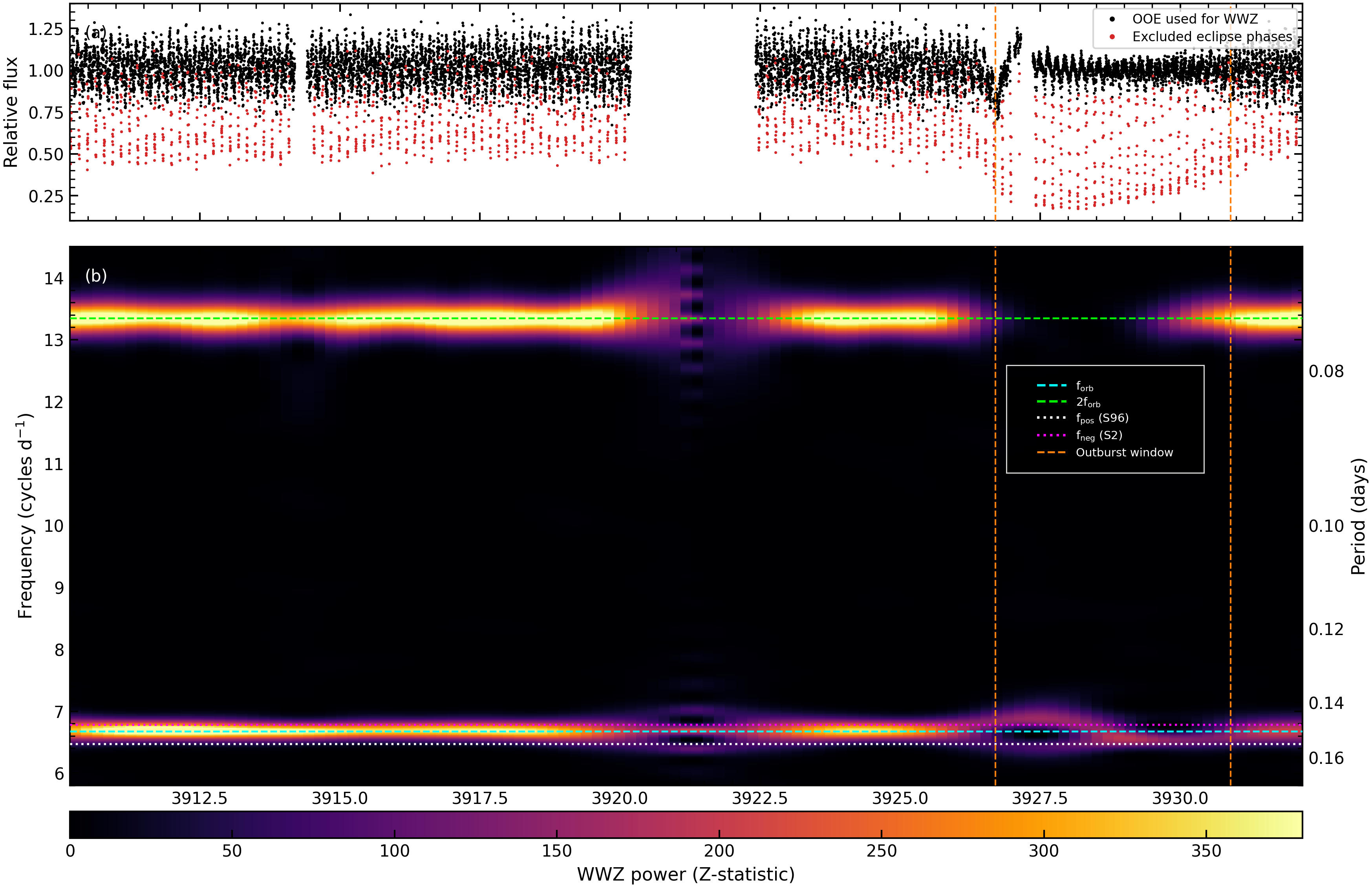}
    \caption{
    WWZ time-frequency map of the eclipse-masked, LOWESS-detrended Sector~96 
    OOE light curve ($c = 0.0003$). 
    \textit{Top panel:} the detrended light curve (black) and eclipse-masked OOE 
    data (red); orange dashed lines mark the outburst boundaries. 
    \textit{Bottom panel:} WWZ power (Z-statistic) as a function of time and 
    frequency. The cyan dashed line marks $f_{\rm orb}$, the green dashed line 
    marks $2f_{\rm orb}$, the white dotted line marks the static LS outburst 
    detection at $P_{\rm LS} = 0.154468$~d, and the magenta dotted line marks 
    the Sector~2 negative superhump candidate at $P_{\rm LS} = 0.147417$~d. 
    Three features are highlighted: the persistent $2f_{\rm orb}$ harmonic 
    during quiescence, its suppression at outburst onset, and the localized 
    enhancement near $f \approx 6.45$--$6.50$~cycles~d$^{-1}$ during the 
    outburst interval consistent with the positive superhump candidate.
    }
    \label{fig:s96_wwz}
\end{figure*}

\subsection{State-dependent orbital variability and superhump-like signals}
\label{subsec:state_dep_orbital_var}

The two \textit{TESS} outbursts of ASASSN-15af differ markedly in duration and morphology. The Sector~2 event lasts approximately 10~d, reaches about $2.2$--$2.3$~mag above quiescence, and shows a precursor-like shoulder before the main bright phase, whereas the Sector~96 event is shorter ($\sim4$~d), lower in amplitude ($\sim1.6$--$1.8$~mag), and more nearly single-peaked (Figures~\ref{fig:tequila_lc} and~\ref{fig:spoc_all_sectors}). The possible physical interpretation of these differences is discussed in Section~\ref{subsec:disc_superhumps}.

To examine how the photometric variability changes with accretion state, we analyzed the quiescent and outburst intervals in Sectors~2 and 96 separately. Each subset was detrended independently and folded on the adopted orbital ephemeris. After excluding the eclipse core, we examined the OOE waveform and its dominant periodicity using Lomb--Scargle \citep[LS;][]{lomb_1976,scargle_1982} periodograms. For Sector~96, we also computed a Weighted Wavelet Z-transform \citep[WWZ;][]{foster_1996} to track the time dependence of the dominant periodicities across the outburst.

\subsubsection{Sector~2}
\label{subsubsec:s2_state_compare}

Figure~\ref{fig:s2_state_compare} compares the quiescent and outburst OOE behavior in Sector~2. In quiescence, the dominant LS peak occurs at $P_{\rm LS}=0.149816$~d, consistent with the orbital period. During outburst, the OOE waveform becomes substantially flatter and the dominant period shifts to $P_{\rm sh,-}=0.14741\ \pm 0.0002$~d, approximately $1.6\%$ shorter than $P_{\rm orb}$. A free-frequency fit is strongly preferred over one fixed at the orbital frequency ($\Delta{\rm BIC}\simeq102$).

Using
\begin{equation}
    \frac{1}{P_{\rm prec}}=
    \frac{1}{P_{\rm sh, -}}-\frac{1}{P_{\rm orb}},
    \label{eq:beat_negative}
\end{equation}
the corresponding beat period is $P_{\rm prec}\approx9.2$~d. Because this is comparable to the $\sim10$~d duration of the outburst, the value should be regarded as approximate.

A time-resolved WWZ analysis was not pursued for Sector~2 because the 30-minute FFI cadence provides only $\sim7$ samples per orbital cycle, limiting our ability to track the detailed evolution of the orbital variability during the outburst.

\subsubsection{Sector~96}
\label{subsubsec:s96_state_compare_period}

Figure~\ref{fig:s96_state_compare} shows the corresponding comparison for Sector~96. In quiescence, the OOE waveform is strongly double-humped, and the dominant LS peak occurs at $P_{\rm LS}=0.074914$~d, close to $P_{\rm orb}/2$. During outburst, the dominant signal shifts to $P_{\rm sh, +}=0.1545\pm0.0006$~d, approximately $3.1\%$ longer than the orbital period. A free-frequency fit is strongly preferred over one fixed at the orbital frequency ($\Delta{\rm BIC}\simeq265$)

The corresponding beat relation,
\begin{equation}
    \frac{1}{P_{\rm prec}}=
    \frac{1}{P_{\rm orb}}-\frac{1}{P_{\rm sh,+}},
    \label{eq:beat_positive}
\end{equation}
gives $P_{\rm prec}\approx5.0$~d.

The WWZ map in Figure~\ref{fig:s96_wwz} confirms that the dominant periodic behavior changes during the outburst. In quiescence, power is concentrated near $2f_{\rm orb}\approx13.35$~cycles~d$^{-1}$, corresponding to the persistent double-humped OOE waveform. This component weakens at the onset of the outburst, while power appears near $6.45$--$6.50$~cycles~d$^{-1}$ ($P\approx0.154$--$0.155$~d), coincident with the longer-period LS signal.

Thus, the two outbursts show periodic signals displaced in opposite directions from the orbital period: the Sector~2 signal is $\sim1.6\%$ shorter than $P_{\rm orb}$, whereas the Sector~96 signal is $\sim3.1\%$ longer. Their physical interpretation is considered in Section~\ref{subsec:disc_superhumps}.

\subsection{State-dependent eclipse morphology}
\label{subsec:results_state_compare}

Sector~96 offers the clearest opportunity to compare the eclipse morphology between quiescence and outburst within the same \textit{TESS} sector. Although Sector~2 also contains both states, its 30-minute FFI cadence significantly smears the eclipse profile and limits sensitivity to changes in eclipse depth, centroid, and width. We therefore restrict the quantitative state comparison to the 2-minute Sector~96 data.

We characterized the eclipse profiles using weighted phase moments within $|\phi|<0.07$ and duration measurements from the median-binned light curves. Uncertainties were estimated from 2000 bootstrap resamples of the unbinned phase--flux data, with the baseline normalization repeated for each realization. The resulting measurements are summarized in Table~\ref{tab:eclipse_duration}.

The dominant state-dependent changes are a substantial increase in eclipse depth and a shift in the flux-weighted eclipse centroid. During outburst, the eclipse depth increases by $\Delta D=0.196\pm0.028$, while the centroid shifts earlier by $\Delta t_{\rm c}=-66.6\pm7.3$~s. By contrast, the eclipse-core width and asymmetry change only weakly within the bootstrap uncertainties.

Thus, the outburst is characterized primarily by a deeper eclipse and a shift in the flux-weighted eclipse centroid, rather than by a major change in the eclipse-core width. The increased fractional eclipse depth is naturally explained if the accretion disk becomes substantially brighter during outburst while the donor-star contribution remains approximately unchanged, increasing the fraction of the system light removed during eclipse. The centroid shift further indicates that the spatial distribution of the eclipsed accretion light changes between quiescence and outburst.

The quiescent eclipse morphology is otherwise broadly stable across sectors. The Sector~29 profile and the quiescent portion of Sector~96 have comparable depths and characteristic widths within their bootstrap uncertainties, although their profile centroids differ by approximately 18~s. Because we measure these centroids from the flux-weighted folded eclipse profiles, this small offset is more naturally attributed to changes in the asymmetric accretion-light distribution than to a change in the orbital ephemeris.

\begin{deluxetable*}{lcccc}
\tablecaption{Bootstrap Eclipse Measurements
\label{tab:eclipse_duration}}
\tablehead{
\colhead{State} &
\colhead{Depth} &
\colhead{FWHM (min)} &
\colhead{Width at 10\% depth (min)} &
\colhead{Centroid (s)}
}
\startdata
Sector~29            & $0.5100^{+0.0087}_{-0.0053}$ & $19.15^{+0.26}_{-0.26}$ & $32.77^{+0.50}_{-0.25}$ & $+76.58^{+2.80}_{-2.96}$ \\
Sector~96 quiescence & $0.5463^{+0.0068}_{-0.0096}$ & $19.96^{+0.69}_{-0.45}$ & $39.19^{+1.38}_{-4.51}$ & $+58.71^{+3.81}_{-3.98}$ \\
Sector~96 outburst   & $0.7426^{+0.0320}_{-0.0208}$ & $19.28^{+0.46}_{-0.52}$ & $32.49^{+1.65}_{-1.72}$ & $-7.87^{+5.98}_{-6.34}$ \\
\hline
Outburst $-$ quiescence &
$+0.1963 \pm 0.0276$ &
$-0.69 \pm 0.75$ &
$-6.69 \pm 3.39$ &
$-66.58 \pm 7.29$ \\
\enddata
\tablecomments{
Depths, widths, and centroids measured from median-binned phase-folded profiles after application of the quality mask. Uncertainties are $1\sigma$ bootstrap intervals from 2000 resamples of the unbinned phase--flux data, with the out-of-eclipse baseline centering repeated independently for each realization.
}
\end{deluxetable*}

\subsection{Broadband SED and Gaia CMD position}
\label{subsec:results_sed}

We constructed a broadband spectral-energy distribution of ASASSN-15af using the archival Johnson $B,V$, Gaia~$G$, 2MASS~$JHK_s$, and WISE~$W1,W2$ photometry listed in Table~\ref{tab:archival_phot}, and fitted the data with the \texttt{SEDFit} code \citep{kounkel_2023}, adopting the Gaia parallax distance of $\sim392$~pc.

A simple two-component model provides an adequate description of the eight-band SED. The fit consists of a cool component with $T_{\rm eff}\simeq3380$~K and $R\simeq0.42\,R_\odot$, together with a hotter component with $T_{\rm eff}\simeq1.4\times10^4$~K and $R\simeq0.026\,R_\odot$. We interpret the cool component as broadly representative of the donor contribution and the hot component as accretion-dominated emission associated with the white dwarf and inner accretion flow. Given the sparse, non-simultaneous photometry and the intrinsic variability of the source, these parameters should be regarded as a phenomenological description. The resulting two-component fit is shown in Figure~\ref{fig:sed_2comp}.

We also placed ASASSN-15af on the Gaia color--absolute-magnitude diagram of known cataclysmic variables from \citet{abrahams_2022}. Using the Gaia DR3 mean magnitude $G=16.9918$, color $G_{\rm BP}-G_{\rm RP}=1.2032$, and parallax distance $d=391.6$~pc, we obtain $M_G=9.03$ before extinction correction. ASASSN-15af lies within the locus occupied by known CVs with $2\leq P_{\rm orb}\leq6$~hr (Figure~\ref{fig:gaia_cmd}), including systems with orbital periods comparable to its $P_{\rm orb}=3.596$~hr. Its position in the CMD therefore provides useful population context for the system and is consistent with its identification as an above-gap dwarf nova.

\begin{figure}
    \centering
    \includegraphics[width=0.48\textwidth]{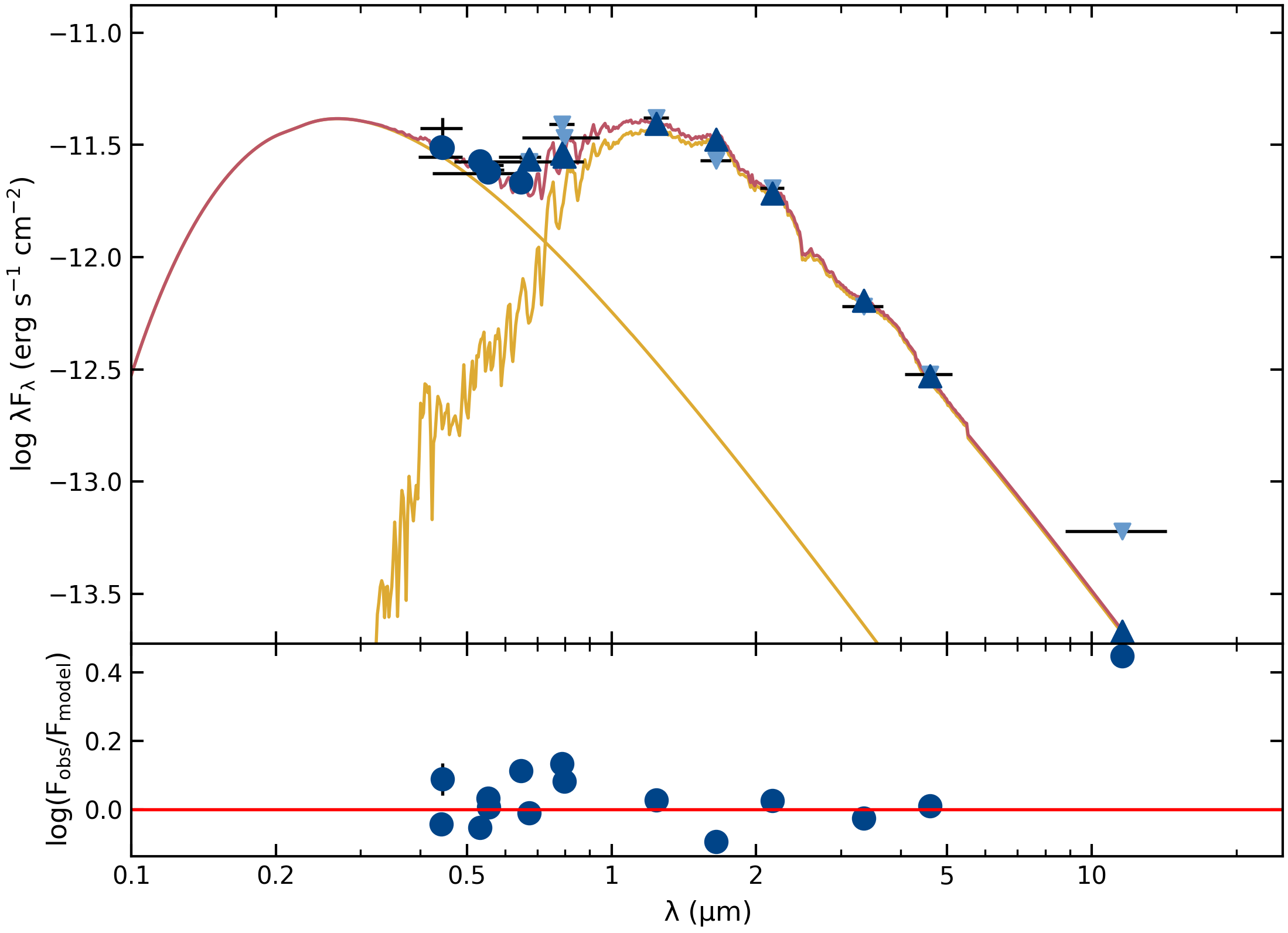}
    \caption{
    Broadband spectral-energy distribution of ASASSN-15af fitted with a two-component phenomenological model using \texttt{SEDFit} \citep{kounkel_2023}. The fit uses Johnson $B,V$, Gaia~$G$, 2MASS~$JHK_s$, and WISE $W1,W2$ photometry at the Gaia distance of $\simeq392$~pc. The cool component ($T_{\rm eff}\simeq3380$~K, $R\simeq0.42\,R_\odot$) is consistent with a donor contribution, while the hot component ($T_{\rm eff}\simeq1.4\times10^{4}$~K, $R\simeq0.026\,R_\odot$) represents the combined hot accretion-related emission.
    }
    
    \label{fig:sed_2comp}
\end{figure}

\begin{figure}[t!]
    \centering
    \includegraphics[width=0.75\linewidth]{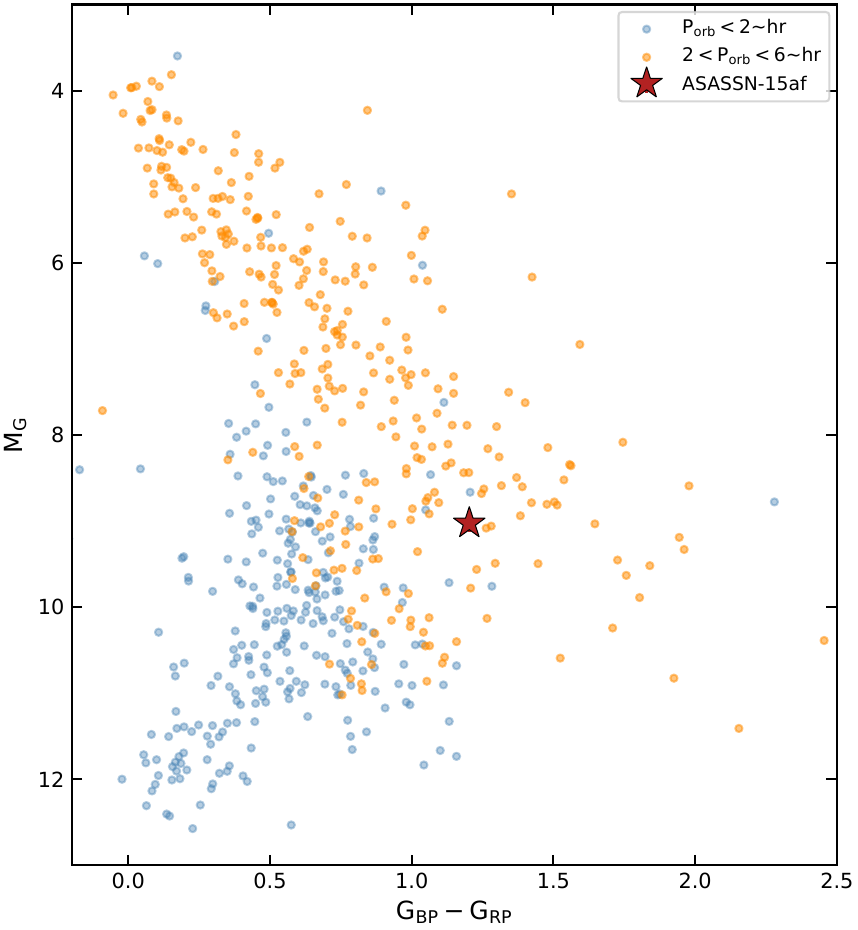}

    \caption{
    Gaia color--absolute-magnitude diagram of ASASSN-15af compared with the cataclysmic-variable sample of \citet{abrahams_2022}. Blue points denote systems with $P_{\rm orb}<2$~hr and orange points those with $2\leq P_{\rm orb}\leq6$~hr. ASASSN-15af is marked by the red star at $G_{\rm BP}-G_{\rm RP}=1.2032$ and $M_G=9.03$, and lies within the locus occupied by known above-gap CVs.
    }
    \label{fig:gaia_cmd}
\end{figure}





\section{Discussion}
\label{sec:discussion}

The combined \textit{TESS}, ASAS-SN, and ATLAS photometry establishes ASASSN-15af as a high-inclination eclipsing dwarf nova above the period gap, with $P_{\rm orb}\simeq0.14982$~d (3.596~hr). Its long-term light curve shows recurrent dwarf-nova outbursts, while the \textit{TESS} observations reveal pronounced state-dependent changes in the eclipse profile and out-of-eclipse variability. Most notably, the two observed outbursts show periodic signals displaced to opposite sides of the orbital period, which we interpret as negative- and positive-superhump-like modulations. Because each behavior is detected in only one outburst, their recurrence and long-term coherence remain to be established.

\subsection{Accretion-state changes and eclipse morphology}
\label{subsec:disc_states}

The eclipse morphology changes measurably between quiescence and outburst. In Sector~96, the outburst eclipse is substantially deeper than in quiescence, with $\Delta D=0.196\pm0.028$, and its flux-weighted centroid shifts earlier by $\Delta t_{\rm c}=-66.6\pm7.3$~s, whereas the eclipse-core width and asymmetry change only weakly. The increased fractional depth is naturally explained if the accretion disk brightens while the donor-star contribution remains approximately unchanged, so that a larger fraction of the system light originates in the eclipsed accretion flow. The small change in eclipse width suggests that the dominant effect is a redistribution of luminosity rather than a major change in the underlying binary geometry.

The centroid shift provides further evidence that the distribution of eclipsed accretion light changes with state. In quiescence, the flux-weighted eclipse centroid occurs slightly after phase zero, with offsets of $+58.7$~s in Sector~96 and $+76.6$~s in Sector~29, whereas during the Sector~96 outburst it moves close to zero ($-7.9$~s). In cataclysmic variables, asymmetric emission from structures such as the stream--disk impact region can displace the flux-weighted eclipse profile from the conjunction of the stellar components \citep{horne_1985,wood_1986}. The reduced offset during outburst is therefore consistent with such asymmetric emission becoming less important relative to the brightened accretion disk.

The persistent $2f_{\rm orb}$ component is consistent with a substantial ellipsoidal contribution from the donor, but its semi-amplitude varies between quiescent sectors. This indicates that the observed double-wave modulation is not produced by the binary geometry alone, but instead reflects a donor-related ellipsoidal signal superposed with variable accretion-related emission. The relative stability of the $2f_{\rm orb}$ phase supports the presence of a persistent orbital-geometry component, while the changing amplitude shows that its observed strength is modulated by the accretion-light contribution. This variability also prevents $A_{2f}$ from being converted directly into a unique donor light fraction. Such an estimate depends on the donor contribution together with the mass ratio, inclination, Roche geometry, limb and gravity darkening, irradiation, and the contributions of the disk and bright spot. A quantitative decomposition would therefore require a joint Roche-star--disk--bright-spot model.

Similar state-dependent changes in the relative contributions of the donor, disk, and asymmetric accretion structures have been reported in other eclipsing dwarf novae \citep{groot_1998,biro_2000,baptista_2000_exdraconis,baptista_2001,scaringi_2013,court_2019,joshi_2024,han_2026}.

\begin{table*}
\centering
\caption{Mass-ratio estimates and Roche-lobe consistency checks for ASASSN-15af.}
\label{tab:mass_ratios}
\begin{tabular}{lllllll}
\hline
Method & Input & Calibration & $q$ & $M_2$ & $R_{\rm L,2}$ \\
       &       &             &     & ($M_\odot$) & ($R_\odot$) \\
\hline

Positive-superhump excess
    & $\epsilon_{+}=0.031$
    & \citet{patterson_2005}
    & $0.14\pm0.04$
    & $0.11\pm0.03$
    & $0.26$ \\

Negative-superhump deficit
    & $|\epsilon_{-}|=0.016$
    & \citet{wood_2009}, eq.~11
    & $0.060$
    & $0.048$
    & $0.20$ \\

Period-offset ratio
    & $\phi=|\epsilon_{-}/\epsilon_{+}|=0.51$
    & \citet{wood_2009}, eq.~17
    & $0.24$
    & $0.19$
    & $0.31$\\

Donor sequence
    & $P_{\rm orb}=3.596$~hr
    & \citet{knigge_2011}
    & ${\sim}0.30$
    & ${\sim}0.24$
    & $0.33$ \\

\hline
\end{tabular}
\tablecomments{
The secondary masses assume $M_1=0.8\,M_\odot$. Roche-lobe radii were
computed using the Eggleton approximation at the observed orbital period,
$P_{\rm orb}=3.596$~hr.
 }
\end{table*}

\subsection{Outburst morphology and superhump-like behavior}
\label{subsec:disc_superhumps}

\subsubsection{Outburst classification and disk-precession states}
\label{subsubsec:outburst_classification}

The contrasting morphologies of the two \textit{TESS} outbursts suggest that they may represent different outburst states. The longer Sector~2 event, with its precursor-like shoulder, resembles a superoutburst, whereas the shorter and simpler Sector~96 event is more consistent with a normal dwarf-nova outburst. This distinction is unusual for a system with $P_{\rm orb}=3.596$~hr because classical SU~UMa dwarf novae are concentrated below the period gap. Above-gap systems showing superoutburst-like behavior are nevertheless known; a relevant example is ASASSN-18aan, an eclipsing SU~UMa-type dwarf nova with $P_{\rm orb}=3.59$~hr \citep{wakamatsu_2021}. The Sector~2 morphology alone therefore does not establish an SU~UMa classification, but it motivates comparison with systems capable of developing an eccentric, precessing disk at relatively long orbital periods.

The period displacements measured during the two outbursts provide further evidence for changing disk states. The shorter-than-orbital signal in Sector~2, with $\epsilon_-\simeq-0.016$, is consistent with a negative superhump, commonly associated with retrograde nodal precession of a tilted disk. By contrast, the longer-than-orbital signal in Sector~96, with $\epsilon_+\simeq+0.031$, is consistent with a positive superhump produced by prograde apsidal precession of an eccentric disk \citep{whitehurst_1988,whitehurst_king_1991,harvey_1995,montgomery_2009,wood_2009}. This ordering is noteworthy because, in classical SU~UMa systems, positive superhumps are usually associated with superoutbursts, which usually take several days to develop. In ASASSN-15af, however, the positive-superhump-like signal is seen in the shorter (normal) outburst, while the longer event instead shows the negative-superhump-like modulation. Given that each signal is detected in only one outburst, it remains unclear whether this contrast reflects a persistent physical difference between the two outburst states

The ratio of the two fractional period offsets,
\[
\phi=\left|\frac{\epsilon_-}{\epsilon_+}\right|\simeq0.51,
\]
is close to the empirical tendency for the negative-superhump deficit to be approximately half the positive-superhump excess in systems showing both modes \citep{patterson_1999}. This agreement supports a disk-precession interpretation, although the two signals were observed in separate outbursts and not simultaneously.

The Sector~96 WWZ analysis provides no evidence for a simultaneous enhancement at the Sector~2 negative-superhump-like frequency during the positive-superhump-like episode. The present observations therefore favor an epoch-dependent picture in which different disk configurations dominate different outbursts instead of requiring the two modes to coexist. Additional high-cadence observations during future outbursts will be needed to determine whether either modulation recurs and whether transitions between the two states are reproducible.

\subsubsection{Mass-ratio constraints from the superhump periods}
\label{subsubsec:superhump_mass_ratio}

If the displaced outburst-state periods are interpreted as superhumps, they provide model-dependent estimates of the binary mass ratio. For the positive-superhump-like signal, $\epsilon_+\simeq0.031$, the empirical relation of \citet{patterson_2005},
\[
\epsilon_+ \simeq 0.18q+0.29q^2,
\]
gives $q\simeq0.14\pm0.04$. For the negative-superhump-like signal, $\epsilon_-\simeq-0.016$, the numerical calibration of \citet{wood_2009} gives $q\simeq0.06$, while applying the $\phi$-based calibration gives $q\simeq0.24$. For comparison, the semi-empirical donor sequence at $P_{\rm orb}\simeq3.6$~hr implies $q\simeq0.30$ for a representative white-dwarf mass \citep{knigge_2011}. These estimates are summarized in Table~\ref{tab:mass_ratios}.

The wide range of inferred values reflects the limitations of applying these calibrations to a system in which the two candidate superhump modes occur in separate outbursts and their evolutionary stages are unknown. In particular, the Sector~96 event is too short to identify a stage~A--B--C sequence, while the negative-superhump relation depends on disk radius, tilt, and surface-density structure. The $\phi$-based estimate lies closer to the donor-sequence expectation, but it additionally assumes that period offsets measured in different disk states can be combined meaningfully.

We therefore do not adopt a unique photometric mass ratio. The available estimates indicate only a broad range of approximately $q\simeq0.1$--$0.3$, and time-resolved spectroscopy will be required for a secure determination.

\subsection{O--C curvature and long-term evolution}
\label{subsec:disc_pdot}

The eclipse timings show a formal preference for a quadratic ephemeris, corresponding to
\[
\frac{dP}{dE}=(2.10\pm0.63)\times10^{-10}\ {\rm d~cycle^{-1}}
\]
and an apparent
\[
\dot{P}_{\rm orb}\simeq1.40\times10^{-9}\ {\rm s~s^{-1}}
\]
($\simeq0.044$~s~yr$^{-1}$). However, the timings are concentrated in only three widely separated \textit{TESS} epochs and require an additional timing-jitter term of $\simeq27$~s, so the curvature should not be interpreted as a secure measurement of secular orbital evolution.

If taken literally as conservative mass transfer, the apparent $\dot{P}_{\rm orb}$ would imply a mass-transfer rate of order $10^{-6}\,M_\odot\,{\rm yr}^{-1}$ for a representative above-gap donor, approximately three orders of magnitude larger than expected for magnetic-braking-driven CV evolution \citep{knigge_2011}. A secular period increase of this magnitude is therefore physically implausible.

Accretion-related timing variability provides a more immediate alternative. In Sector~96, the flux-weighted eclipse centroid shifts by $-66.6\pm7.3$~s between quiescence and outburst (Section~\ref{subsec:results_state_compare}), demonstrating directly that changes in the accretion-light distribution can displace the measured eclipse profile by tens of seconds. Smaller state- or epoch-dependent changes of this kind could mimic long-term curvature when sampled at only a few separated epochs. Cyclic donor-star activity \citep{applegate_1992,lanza_1998} or a light-travel-time effect from a third body could also contribute, but the present baseline is insufficient to test these possibilities.

We therefore regard the quadratic term as tentative evidence for O--C curvature instead of a measurement of secular $\dot{P}_{\rm orb}$. Additional high-cadence eclipse timings over multiple years and accretion states will be required to determine whether the trend persists.

\subsection{Context among above-gap eclipsing CVs}
\label{subsec:disc_context}

With $P_{\rm orb}\simeq3.596$~hr, ASASSN-15af lies above the canonical CV period gap \citep{warner_1995,knigge_2011}. Its deep eclipses and recurrent outbursts support its classification as a high-inclination, above-gap dwarf nova. The detection of both positive and negative superhump-like signals further suggests complex accretion-disc behaviour.

Although its orbital period overlaps the 3--4~hr range commonly associated with SW~Sex systems \citep{rodriguez-gil_2007b,schmidtobreick_2013}, its recurrent dwarf-nova outbursts distinguish it from the predominantly high-state behaviour of classical SW~Sex stars. Moreover, SW~Sex classification relies primarily on spectroscopic diagnostics that are not available for ASASSN-15af. Phase-resolved spectroscopy would therefore be needed to investigate any possible association with this subclass.

\subsection{Limitations and future follow-up}
\label{subsec:disc_limits}

The main limitations of the present study are the available cadence, the limited number of observed outbursts, and the sparse long-term eclipse-timing baseline. The 2-minute \textit{TESS} data resolve the eclipses well enough to characterize state-dependent morphology, but not the individual ingress and egress components required for a full geometric decomposition of the white dwarf, disk, bright spot, and donor. Sector~2 is further limited by its 30-minute cadence. In addition, the positive- and negative-superhump-like signals are each detected in only one outburst, while the eclipse timings effectively sample only three widely separated epochs.

High-speed, preferably multi-band, eclipse photometry would provide stronger constraints on the binary geometry and the relative contributions of the donor, disk, and bright spot \citep[e.g.,][]{horne_1985,wood_1986,littlefair_2008,copperwheat_2010,savoury_2011}. A quantitative estimate of the donor light fraction will require a physical light-curve decomposition combining Roche geometry with the accretion disk and bright spot; multi-band eclipse photometry together with spectroscopic constraints on the mass ratio would substantially reduce the degeneracies in such a model. Time-resolved spectroscopy is also important for measuring radial velocities, constraining the component masses, and testing for SW~Sex spectroscopic characteristics.

Continued high-cadence monitoring of future outbursts will establish whether the two superhump-like signals recur and how they evolve with accretion state, while additional eclipse timings over a longer baseline will determine whether the tentative O--C curvature persists or instead reflects accretion-related timing variability.

\section{Conclusions}
\label{sec:conclusion}

We have presented the first detailed photometric characterization of ASASSN-15af (TIC~231050794; 2MASS~J01575499$-$5430367) using \textit{TESS} observations from Sectors~2, 29, 30, 69, and 96, supplemented by long-term ASAS-SN and ATLAS photometry. The combined data establish ASASSN-15af as a high-inclination eclipsing dwarf nova above the canonical period gap, correcting an earlier catalog association with an unrelated young stellar object.

From 391 eclipse timings in Sectors~29, 30, and 69, we derive
\[
P_{\rm orb}=0.14982063(2)~{\rm d}
\]
($3.596$~hr). The timings show excess scatter at the level of tens of seconds, with a fitted timing-jitter term of $\simeq27$~s. Although a quadratic ephemeris gives a formally improved fit, the timings sample only three widely separated epochs; we therefore regard the apparent O--C curvature as tentative and not as evidence for secular orbital evolution.

The quiescent out-of-eclipse waveform is dominated by a persistent $2f_{\rm orb}$ component whose phase remains nearly constant between sectors, consistent with a substantial ellipsoidal contribution from the donor. Its semi-amplitude nevertheless varies from sector to sector, indicating that the observed double-wave modulation is not produced by binary geometry alone but is modified by changing contributions from the accretion disk and other asymmetric structures. During the Sector~96 outburst, the fractional $2f_{\rm orb}$ amplitude decreases by about 66\%, consistent with dilution by the brighter accretion flow.

The eclipse morphology also changes with accretion state. During the Sector~96 outburst, the eclipse deepens by $\Delta D=0.196\pm0.028$ and its flux-weighted centroid shifts earlier by $\Delta t_{\rm c}=-66.6\pm7.3$~s, while the eclipse-core width changes only weakly. These changes indicate a redistribution of accretion-powered light during outburst.

The most unusual behavior is the appearance of outburst-state periodicities on opposite sides of the orbital period. In Sector~2, the dominant signal is $\sim1.6\%$ shorter than $P_{\rm orb}$, whereas in Sector~96 it is $\sim3.1\%$ longer. We interpret these as negative- and positive-superhump-like modulations, respectively. Their fractional period offsets satisfy
\[
\left|\frac{\epsilon_-}{\epsilon_+}\right|\simeq0.51,
\]
consistent with empirical trends in systems showing both modes. Because each signal is detected in only one outburst, however, these identifications remain provisional and their recurrence has yet to be established.

Taken together, the recurrent dwarf-nova outbursts, state-dependent eclipse morphology, changing harmonic structure, and superhump-like signals indicate that ASASSN-15af contains a dynamically evolving accretion disk. We therefore classify it primarily as an above-gap eclipsing dwarf nova.

Future high-speed multi-band photometry will be important for resolving the detailed eclipse structure and constraining the binary geometry, while continued monitoring will test whether the superhump-like signals recur. Time-resolved spectroscopy is needed to measure the component velocities and mass ratio. It can also test whether the system exhibits any spectroscopic characteristics associated with the SW~Sex phenomenon, but the present photometric data do not support an SW~Sex classification.

\section*{Acknowledgments}

This work was supported in part by the Israel Ministry of Innovation, Science, and Technology under Grant No.~0008107 and by the Israel Science Foundation (ISF) under Grant No.~1404/22. This work made use of data from the Mikulski Archive for Space Telescopes (MAST). The specific TESS light curves analyzed in this study are available via \dataset[https://doi.org/10.17909/t9-nmc8-f686]{https://doi.org/10.17909/t9-nmc8-f686}. Funding for the TESS mission is provided by NASA's Explorer Program. The Space Telescope Science Institute (STScI) is operated by the Association of Universities for Research in Astronomy, Inc., under NASA contract NAS~5--26555. B.B.O. acknowledges support from Ariel University through the postgraduate scholarship and research and development funding.

\bibliographystyle{aasjournalv7} 
\bibliography{asass_15af.bib}
\end{document}